\documentclass[11pt]{article}
\usepackage{moriond2000,epsfig}

\def\be{\begin{equation}}
\def\ee{\end{equation}}
\def\bea{\begin{eqnarray}}
\def\eea{\end{eqnarray}}

\begin{document}
\vspace*{4cm}
\title{WEAK WEAK LENSING: HOW ACCURATELY CAN SMALL SHEARS BE MEASURED?}

\author{ K. KUIJKEN}

\address{Kapteyn Instituut, PO Box 800, 9700 AV Groningen, The Netherlands}

\maketitle\abstracts{ Now that weak lensing signals on the order of a
percent are actively being searched for (cosmic shear, galaxy-galaxy
lensing, large radii in clusters...) it is important to investigate
how accurately weak shears can be determined. Many systematic effects
are present, and need to be understood. I show that the Kaiser et
al. technique can leave residual systematic errors at the percent
level (through imperfect PSF anisotropy correction), and present an
alternative technique which is able to recover shears a factor of ten
weaker.
}

\section{Introduction}

Weak lensing has evolved in the last ten years into a quantitative
tool in cosmology. The goal is no longer to demonstrate a convincing
detection of the effect, but to make real measurements of the shear,
and turn these into real measurements of the projected mass density in
a given direction. 

\begin{figure}
\epsfxsize0.45\hsize
\epsfbox{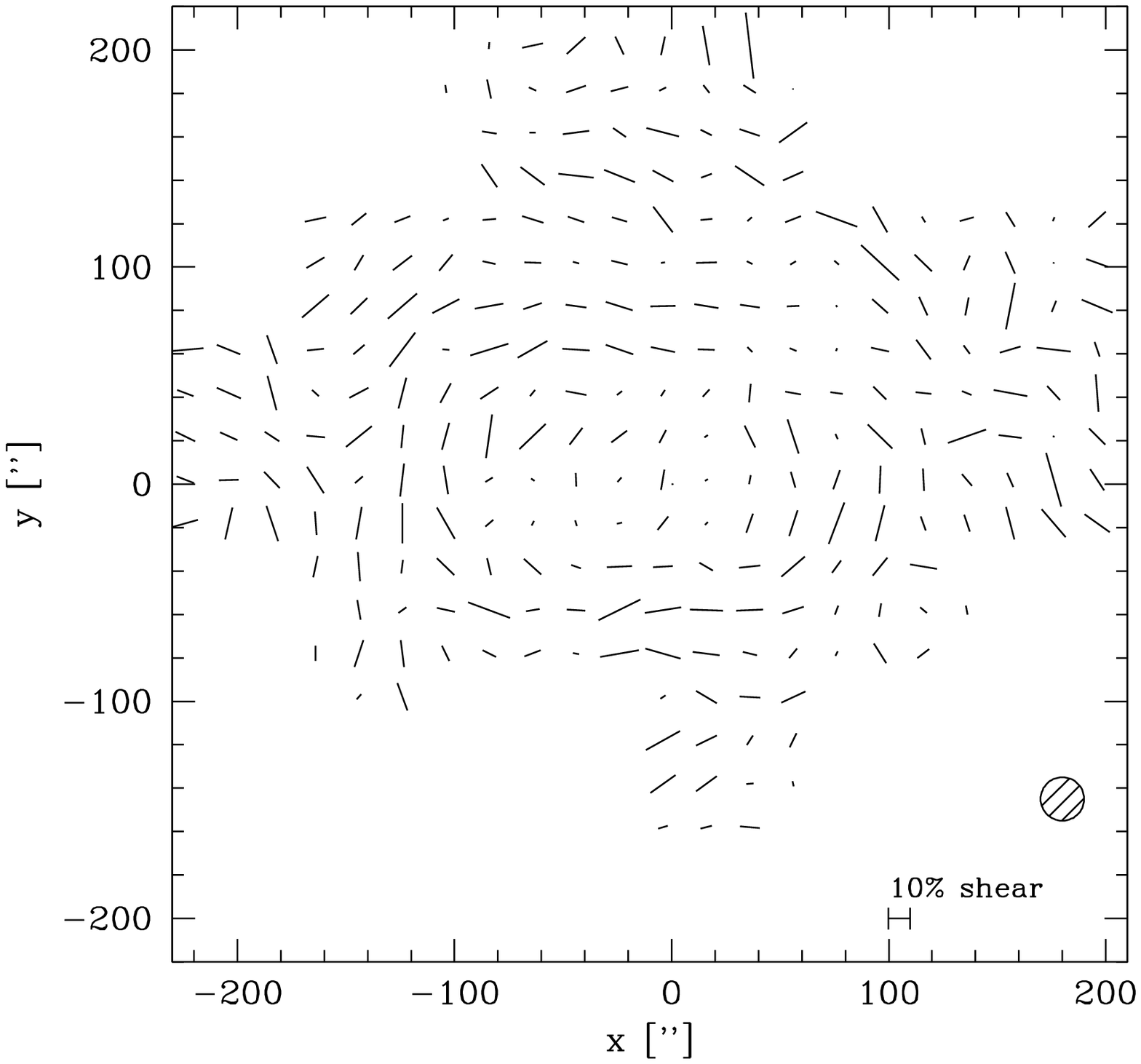}
\epsfxsize0.45\hsize
\epsfbox{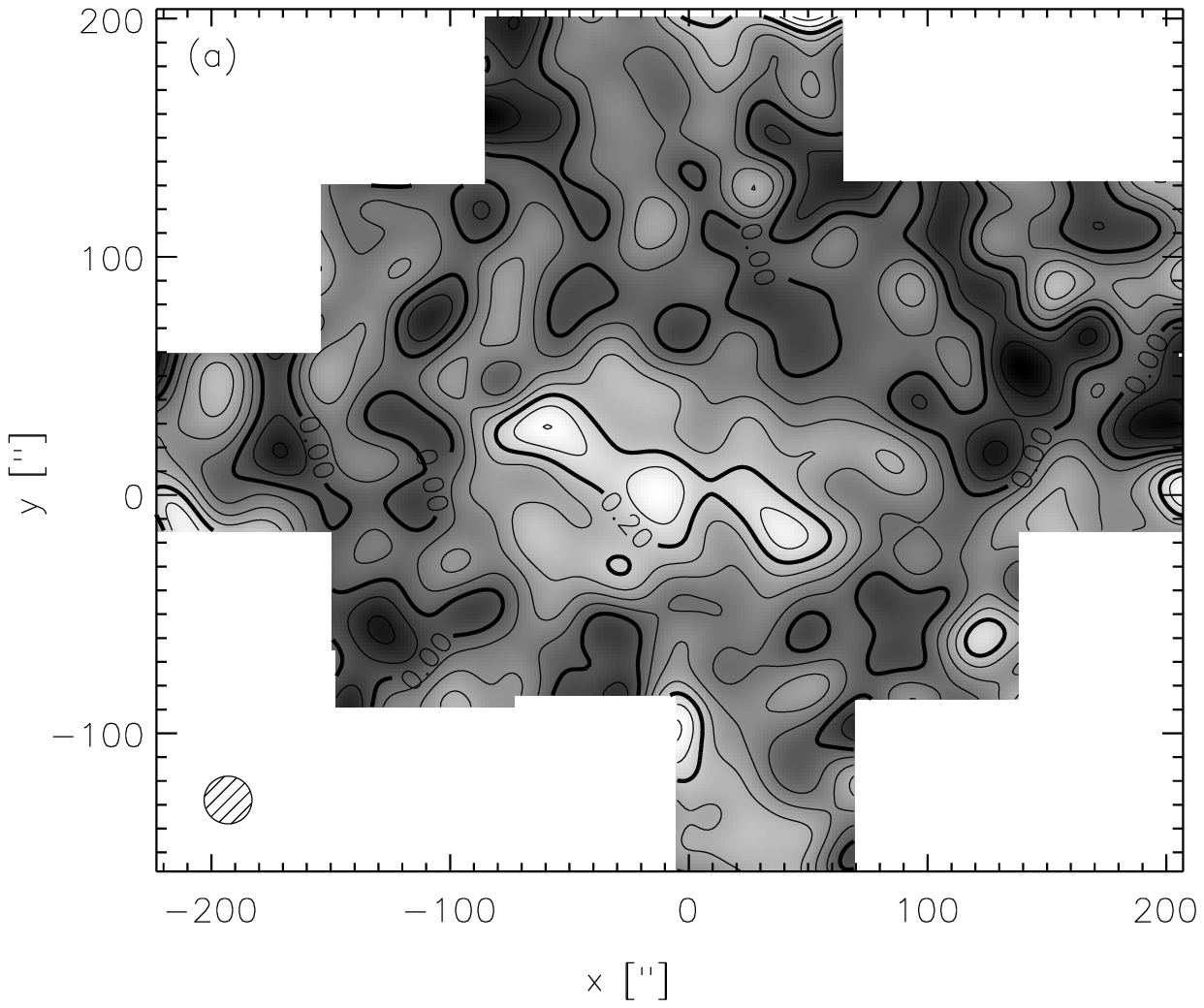}
\caption{Weak Lensing shear field (top), 
and mass map (right) of the $z=0.83$ cluster
MS1054-03 by Hoekstra et al (2000).}
\label{fig:1054}
\end{figure}

An important aspect of this quantitative work is to control systematic
effects. While it is now relatively straightforward to demonstrate
convincingly a coherent alignment of galaxy distortions around a
massive foreground cluster, for example, it is much harder to quantify
the amount of this shear in the presence of seeing, camera
distortions, or uncertain redshift distributions of the source
galaxies. As an example, Fig~\ref{fig:1054} shows the weak shear field
around the $z\sim0.83$ cluster MS1054-03 from a 6-exposure WFPC2
mosaic\cite{2000ApJ...532...88H}. In order to come to this shear field,
many of the effects to be discussed in the following section had to be
quantified and corrected for. 

Nevertheless, this is an example of a rather strong lensing effect,
shears around 10\%.  The field is also evolving in the direction of
working with weaker and weaker gravitational shears. Ground-based
cameras now have wider fields which allow the outskirts of clusters to
be observed efficiently. For a singular isothermal sphere model the
shear falls with radius as $r^{-1}$, and it is now routinely possible
to reach the region of clusters where the shear should be around 1\%.
While formally signal-to-noise is a very weak function of radius (the
shear drops outwards, but the number of available background galaxies
whose shapes may be averaged to yield a shear estimate increases with
radius), systematic effects become a serious concern. Also lower-mass
clusters and groups, much more representative of the universe than the
massive X-ray clusters, are now within reach (Hoekstra et al 2000).

In clusters there is at least the independent sanity check of making
sure that the shear is aligned roughly with the observed
cluster. Lately, though, a lot of effort has gone into the search for
cosmic shear, which is the lensing effect caused by large-scale
structure\cite{2000A&A...358...30V,Baconetal,Wittmanetal,kaiser}). Here
the measurement is rather similar to the early cosmic background
radiation anisotropy experiments, and constitutes the search for an
excess variance. The effect is also on the level of a percent shear or
less, but its geometry is a priory unclear.

Also galaxy-galaxy lensing is now being carried out with enormous
numbers of lens-source pairs, and formal averaging statistics allow
shears of well below a percent to be measured (see Fisher, this
conference). As in galaxy-galaxy lensing the lens and source are
rather close together on the sky, any large-scale systematic
distortion cancels out to first order, making it a little less
susceptible to residual systematics.

\section{Systematic errors}

The image of a distant galaxy that we record on a CCD is a
\begin{itemize}\itemsep=-3pt
\item charge-transferred,
\item pixellated,
\item camera-distorted,
\item atmospherically blurred,
\item gravitationally lensed,
\item random-shape,
\item randomly-oriented
\end{itemize}
galaxy. We can only extract the lensing information if we can control
all these other effects, either by avoiding them, or by measuring and
correcting for their effects.

Fortunately the sky contains a number of calibrators, stars in the
field. These are for our purposes point-like, and so measure the
smearing effects of atmosphere and optics simultaneously with the
galaxies. Galaxy orientations on the sky are random as far as we know. 
CCD effects can be calibrated by observing at different orientations
with respect to the pixel grid, or by using totally different cameras
altogether. The distortion of the camera can be calibrated with
astrometric standard fields, or by comparing offset exposures.

In principle, therefore, it looks as if the required information
exists to disentangle the effects of gravitational lensing from the
other ones. Assuming for the moment that this has been done, this
means that for each background galaxy an estimate for the
distortion can be obtained.

The distortions $g_i$, ($i=1,2$) are equivalent to the axis ratio and
major axis orientation observed when an intrinsically round source is
lensed. $g$ is related to the distortion matrix
\begin{equation}
\left(
\begin{array}{cc}
1-\kappa-\gamma_1 &-\gamma_2\\
-\gamma_2 &1-\kappa+\gamma_1
\end{array}\right)
\end{equation}
through $g_i=\gamma_i/(1-\kappa)$. Here $\gamma$ is the gravitational
shear, and $\kappa$ the convergence. The latter is a measure of the
surface mass density in the lens plane, and is the goal of weak
lensing. 

There are several complications in converting the observed
distortions into a projected mass density. 

\begin{itemize}
\item {\em PSF.} The Point Spread Function affects galaxy images in
various ways. If the PSF is anisotropic, it will imprint this
anisotropy onto the observed image. On the other hand, since it has a
finite width it will tend to increase the size of faint images. As the
PSF acts as a convolution, galaxies of different size are affected by
PSF in different degrees.
\item {\em Camera Shear.} The camera maps the sky onto the focal
plane, but often does not do this without introducing some
distortion. It is easy to show that such a distortion produces shear,
and that this acts as a simple additive effect onto the observed
gravitational shear. 
\item {\em (1-$\kappa$).} The measured distortion is a combination of
$\gamma$ and $\kappa$. To derive the shear it is therefore necessary
to have an estimate of the convergence. This is a fundamental
limitation of weak lensing, and is known as the mass-sheet
degeneracy\cite{1992grle.book.....S}. Usually one assumes that the mass
distribution at the outer edges of the field follows some simple model
(zero, singular isothermal, ...), or one leaves the uncertain $\kappa$
zeropoint in the result\cite{1995ApJ...439L...1K}. 
\item {\em Redshift Distribution.} The deflection angle in lensing
depends on the relative distances between observer, source and
lens. As the source distances are usually unknown, or only known
statistically, the lensing angles, and hence the shears, need to be
corrected to infinite source redshift.  This correction factor is
usually referred to as the $\beta$ factor:
\be
\gamma(\hbox{observed})=\beta \gamma(\hbox{infinite source distance}),
\ee
where $\beta$ is the ratio of lens to source distance. The effect is
most important for distant cluster work, or for situations where
lenses and sources are distributed similarly down the line of sight as
in cosmic shear measurements. Source galaxies used for lensing are
usually so faint that they are beyond the reach of spectroscopic
redshift surveys, so models or photometric redshift studies need to be
used. The analysis performed for MS1054 (fig.~\ref{fig:1054}) relied
on the HDF redshift distributions\cite{fernan,chen}. A discrepancy
of around 10\%, which may well be a form of cosmic variance, exists
between the two Deep Fields.
\item {\em $\beta$ spread.} Even once the mean redshift of the sources
is determined, a second-order effect exists which depends on the width
of the distribution of $\beta$. This is because the distortion is not
linear in $\kappa$, and so the observed distortion is 
\be
g(\hbox{observed})={\gamma(\hbox{observed})\over1-\kappa(\hbox{observed})}
={\beta\gamma\over1-\beta\kappa}=\beta\gamma+\beta^2\kappa\gamma+O(\kappa^2)
\ee
which, when averaged over source galaxies with different redshift
requires knowledge of $\overline{\beta^2}$, and hence of the variance
of the $\beta$-distribution\cite{2000ApJ...532...88H,1997A&A...318..687S}.
\end{itemize}

Most of these effects are important when it comes to properly
calibrating the strength of a detected distortion pattern, and turning
it into a real mass. 

The correction for PSF effects is technically the most difficult of
these steps. The most extensively used method is the
KSB\cite{1995ApJ...449..460K} method, which can be considered the
current `industry standard.' Of the many weak lensing results that
have been obtained to date\cite{1999ARA&A..37..127M}, most have
employed this technique.

\section{Tests of the KSB method}

The KSB method uses a combination of centered second image moments as
its shape statistic. This is a logical choice: these moments measure
the orientation and elongation of an ellipse, which is a reasonable
first approximation to galaxy images. KSB worked out how these image
moments, and the shape statistics derived from them, behave under
various distorting effects. The result is a formalism for deriving
`polarizabilities', matrices which express the response of the image
polarization 
\be
(e_1,e_2)\equiv\left({I_{xx}-I_{yy}\over I_{xx}+I_{yy}},
{2I_{xy}\over I_{xx}+I_{yy}}\right)
\ee
to gravitational shear, and to PSF smearing.  KSB show how these
polarizabilities can themselves be written as combinations of
higher-order image moments. The image moments $I_{ij}$ of the galaxy
image intentity $f(x,y)$ are computed
with a circular weight function $W(r)$, to prevent poisson noise from
dominating the measurements:
\be
I_{ij}=\int dx dy f(x,y) W(r).
\ee

In particular, the effect on $(e_1,e_2)$ of a distant galaxy under
smearing by an anisotropic PSF is given by
\be
\delta e_\alpha=P^{\rm sm}_{\alpha\beta}p_\beta
\ee
where $p$ is the PSF anisotropy, $(I_{xx}-I_{yy}, 2I_{xy})$, this time
constructed from unweighted second moments.

\begin{figure}
\epsfxsize0.45\hsize\epsfbox{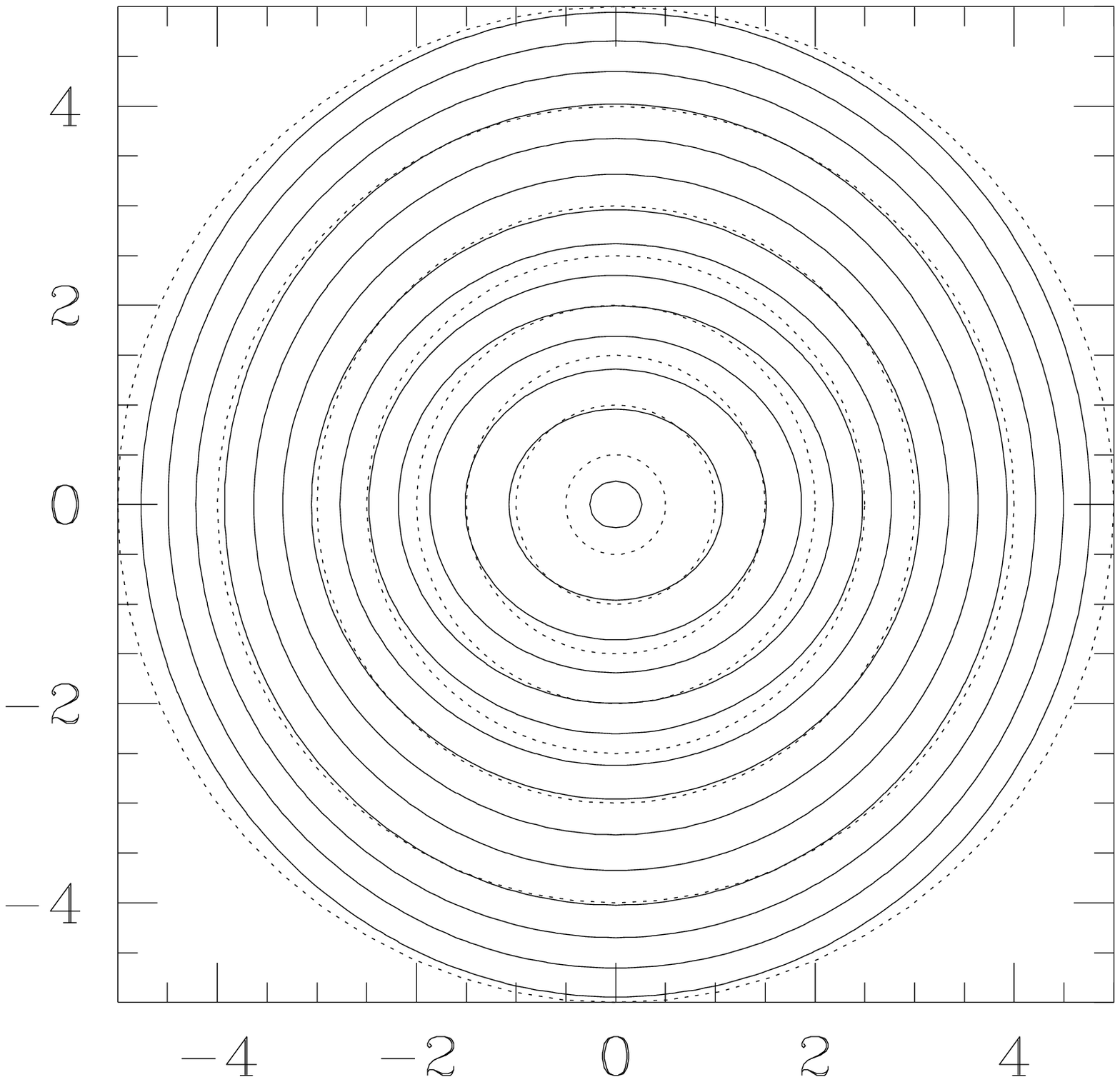}
\epsfxsize0.45\hsize\epsfbox{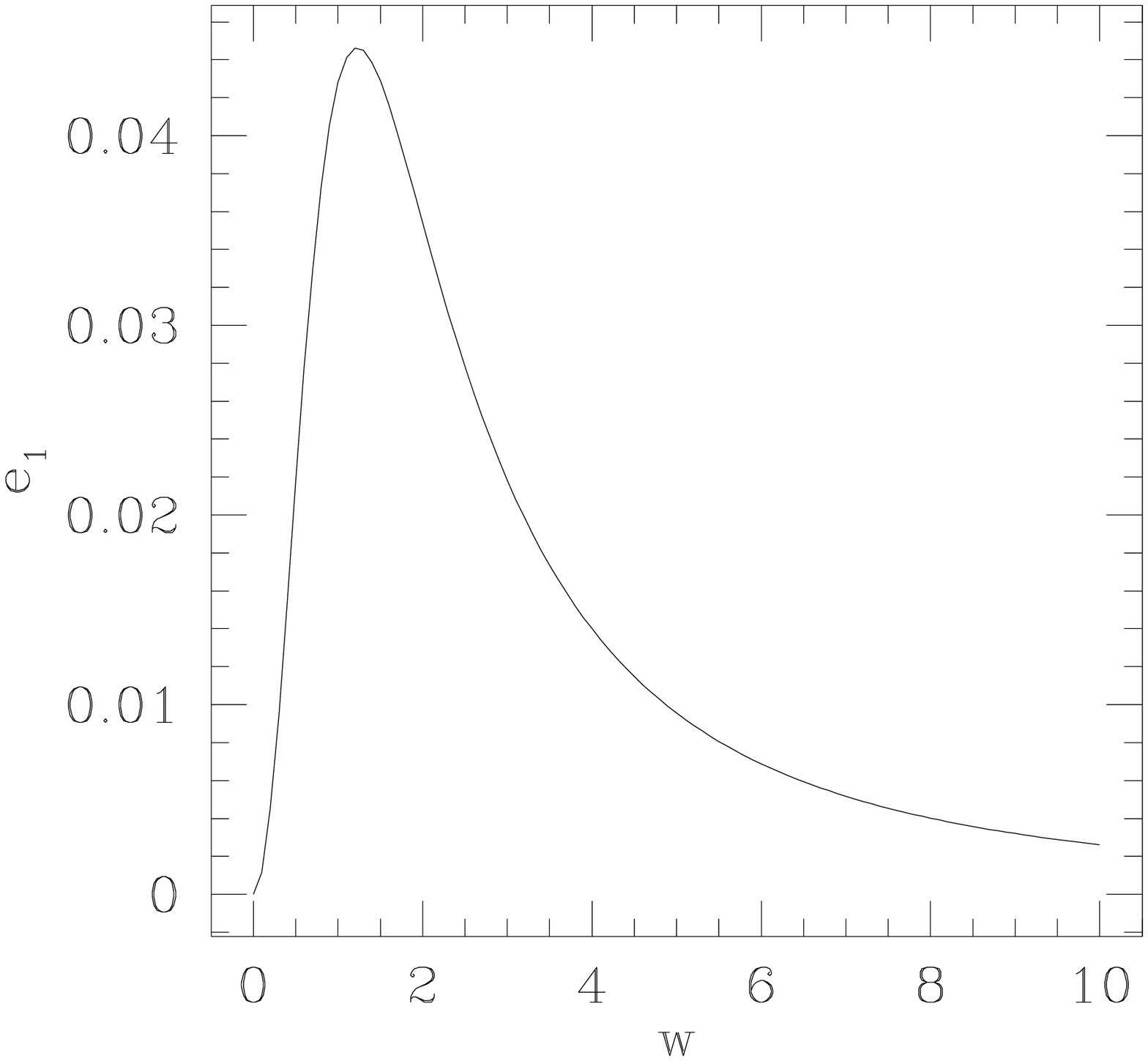}
\caption{Left: An example of a PSF which is isotropic in its second moments,
but not intrinsically. Right: the polarization of this PSF when it is
measured with different gaussian weight functions (1-$\sigma$ radius $w$).}
\label{fig:psf}
\end{figure}

This formalism works very well, but not perfectly. An example is shown
in figure~\ref{fig:psf}, which is the average of two normalized
gaussians, of ($x,y)$-variance $(1.1,1)$ and $(3.9,4)$
respectively. The total variances in $x$ and $y$ of this PSF are the
same, i.e. its anisotropy $p$ is zero. No matter what the
polarizability of a source, therefore, the KSB formulae would predict
no effect on the polarization of a source after convolution with this
PSF. However, as figure~\ref{fig:psf} shows, this is not correct,
because of the radial weight function used in the computation of the
polarizations. Errors of the order of a percent in the image
polarization can result. 

\begin{figure}
\epsfxsize\hsize
\epsfbox{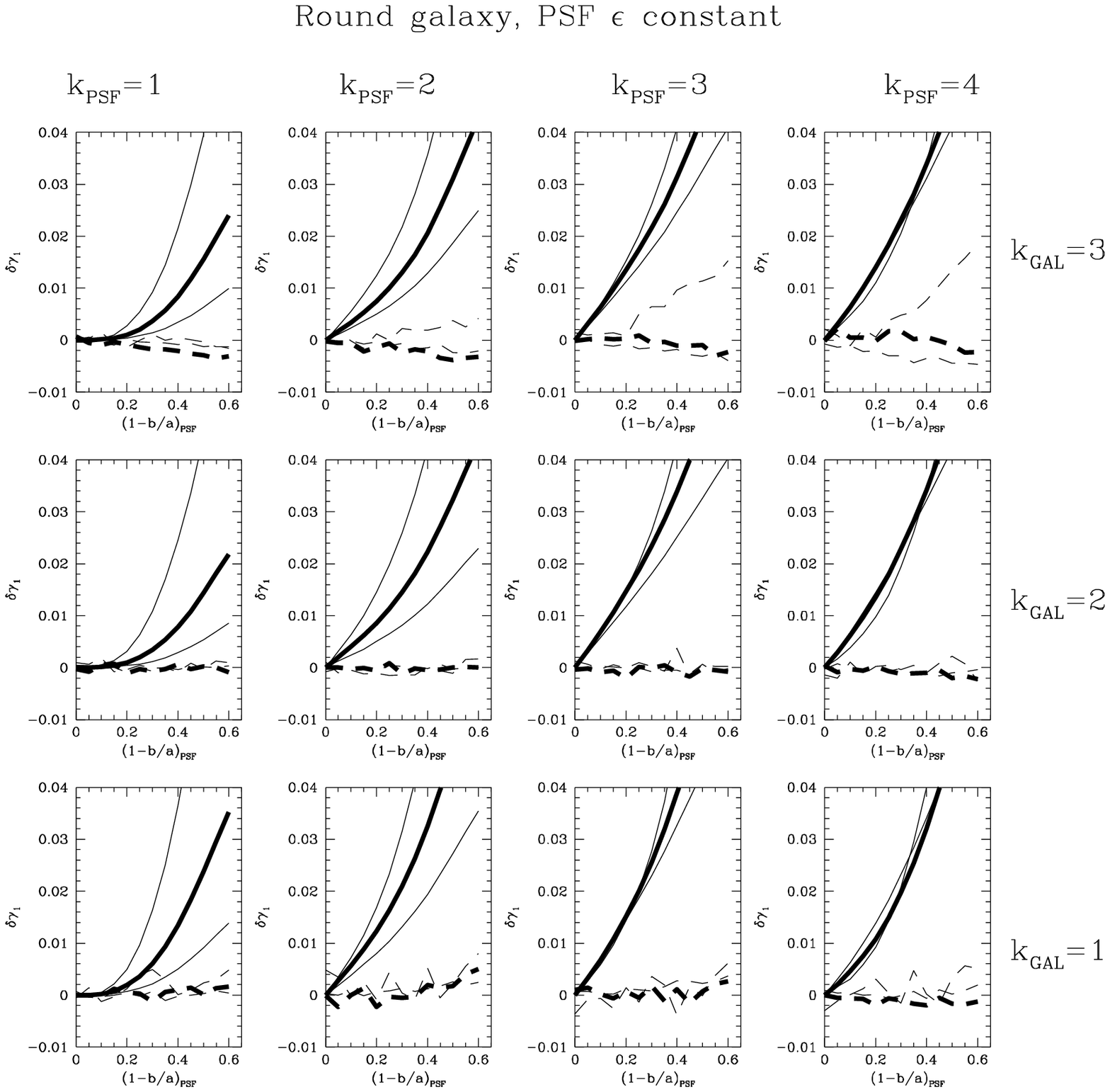}
\caption{Simulations of the extent to which the KSB (solid lines) and
CEO (ashed lines) algorithms corrects an input PSF
anisotropy. Double-gaussian PSF's of axis ratio $b/a$ were convolved
with round double-gaussian `galaxies', and analysed with both
algorithms. $k$ is the ratio of the dispersions of the two gaussian
components: $k=2$ is roughly exponential, $k=3$ is roughly de
vaucouleurs.  No lensing was simulated here, so the derived shears
should be zero. Several percent residuals remain for the most
non-gaussian PSF's.}
\label{fig:ksb}
\end{figure}

A set of simulations illustrating such systematic effects is shown in
fig.~\ref{fig:ksb}. A more extensive discussion is given in
Kuijken~(1999)\cite{1999A&A...352..355K}.

The residual systematics are not fundamental: the PSF is known
perfectly, so all required information is there. It is instead a
consequence of a mathematical assumption made by KSB (that the PSF can
be written as a convolution of a very compact anisotropic kernel with
a more extended, round function) to enable the polarizabilities to be
constructed.

\section{An alternative algorithm: Constant Ellipticity Objects}

Alternative algorithms have been developed, but none are in as wide a
use as KSB. One class is based on reconvolving the data with a
circularizing kernel\cite{Wittmanetal,kaiser2000}. This is a
rather direct way to reduce PSF effects, but creates correlated noise
and, unavoidably, somewhat degrades the data.

An alternative algorithm\cite{1999A&A...352..355K} does not rely on
modelling the second moments, but instead is a direct fit of the
sources to a sheared, instrinsically circular source, convolved with
the PSF. As a sheared circular source has ellipticity which is
constant with radius, the algorithm has been dubbed Constant
Ellipticity Objects (CEO).

\begin{figure}
\epsfxsize\hsize
\epsfbox{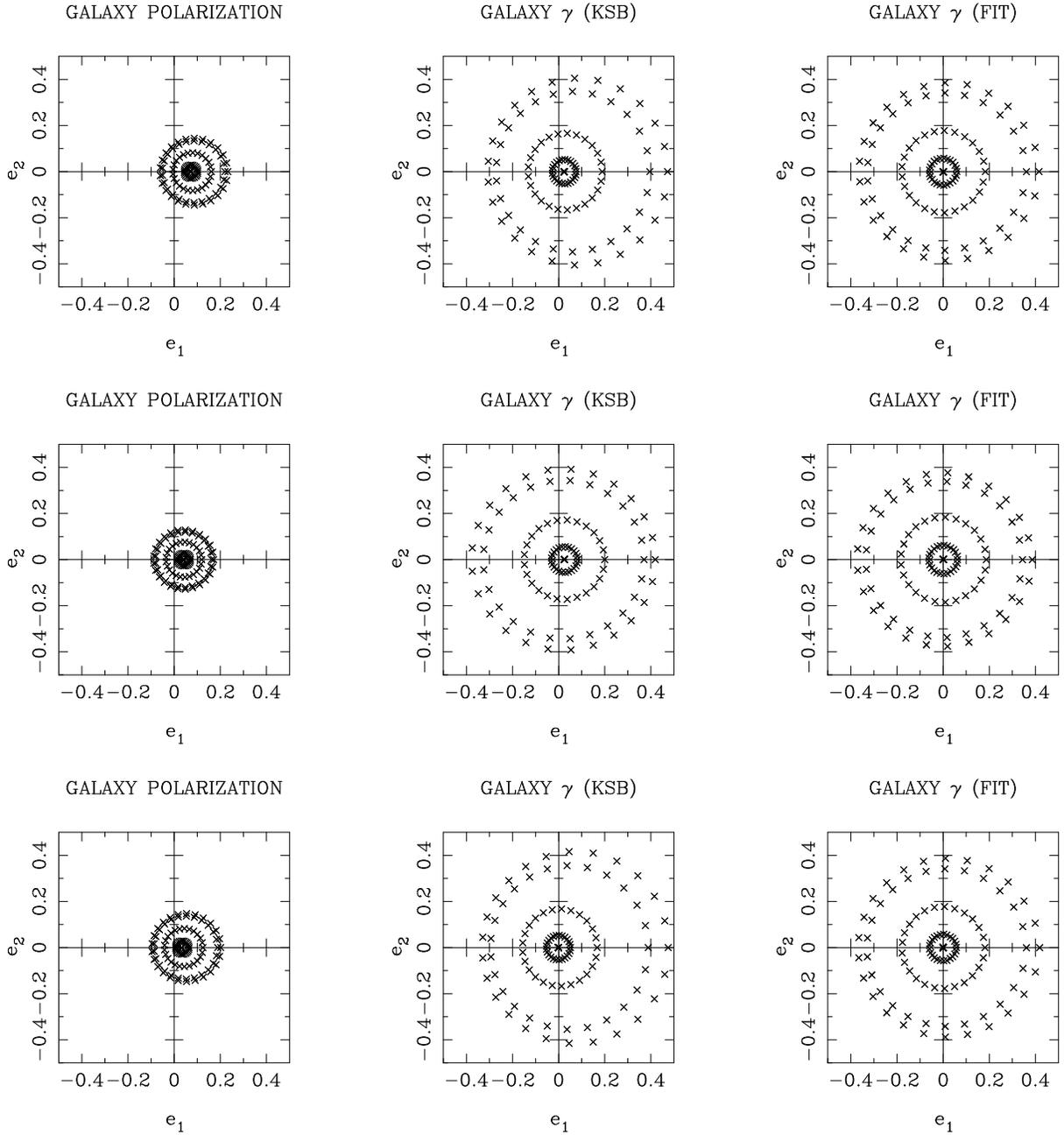}
\caption{Various galaxy models, of different elongations and
orientations, convolved with anisotropic PSFs and then analysed with
the CEO and KSB algorithms. Left column: raw polarizations. Middle
column: shear deduced from the KSB formalism. Right column: shear
deduced from the CEO method. The different rows correspond to
different radial ellipticity profiles of the PSF. }
\label{fig:elong}
\end{figure}

An exhaustive discussion will not be given here, as the details may be
found in the original paper. The essence of the results are as follows:
\begin{itemize}
\item When the sources are instrinsically round, the algorithm
recovers the shears from noise-free images, even for very anisotropic
PSF's which the KSB technique does not correct to better than a percent.
\item When Poisson noise is added to simulated images, the best-fit
shears are unbiased, and the scatter is very similar to the KSB
method. (This shows that the KSB method, even though it only uses a
few moments of each image, is close to optimal in the
amount of information it uses from the images).
\item When the algorithm is applied to sources which are not
intrinsically round, but which are made to look like disk/bulge
systems seen at various orientations and inclinations, the ensemble
average of the individual shear estimates is within a few tenths of a
percent of the correct value. I.e., even though the individual
galaxies are fit with a model which is not correct (they are not
intrinsically round) the errors made average out (see
Fig~\ref{fig:elong}).
\end{itemize}

\section{Consistency Checks}

However shears are measured, it is important to be able to perform
consistency checks on the results. Below are listed a set of tests
that can be performed.
\begin{itemize}\itemsep=-3pt
\item Transform corrected galaxy polarizations $e_1\to e_2$;
$e_2\to-e_1$. Surface mass density should now be consistent with zero.
\item There should be no correlations between $\gamma$ and $e^\star$,
the polarization of the stars used in the correction for PSF effects.
\item Results should be independent of wavelength observed in, or
instrument.
\item Shear fields should not rotate with the instrument or detectors.
\item Smear image data with a typical PSF, and re-analyse these
images. Results may be a little noisier, but should be consistent with
original result. [this tests algorithm, not data]
\item Vary the weight function radius in KSB
\item Track the signal as a function of source size. Smaller sources
are more sensitive to PSF
\end{itemize}

\section{Summary}

Weak lensing work is moving into the regime of very weak
signals. Lensing by large-scale structure, the outskirts of clusters,
and low-mass galaxy groups and individual galaxies are all being
targetted. Particularly the first results being reported elsewhere at
this conference on the detection of lensing by large-scale structure
are very exciting.

Measuring these very weak distortions is a tricky business, because
many other effects need to be characterized and corrected for. I have
described a new algorithm, which performs very well on test data,
which is able to reduce systematic uncertainties associated with the
correction for PSF anisotropy considerably. Applications on real data
are in progress.

\section*{Acknowledgments}
This work, and this conference visit, was supported by the TMR Network
``Gravitational Lensing: New Constraints on Cosmology and the
Distribution of Dark Matter'' of the EC under contract No.\
ERBFMRX-CT97-0172.

\end{document}